\documentclass[prl,twocolumn,superscriptaddress,showpacs,aps]{revtex4-1}

\usepackage{color}
\usepackage{graphicx}
\usepackage{bm}
\usepackage{amsmath}

\begin{document}

\newcommand{\MgB}{MgB$_2$}
\newcommand{\Tc}{T_{\text{c}}}
\newcommand{\Hci}{H_{\text{c1}}}
\newcommand{\Hcii}{H_{\text{c2}}}
\newcommand{\fq}{\phi_0}
\newcommand{\lamn}{\lambda_n}

\title{Metastable Vortex Lattice Phases in Superconducting {\MgB}}

\author{P. Das}
\author{C. Rastovski}
\affiliation{Department of Physics, University of Notre Dame, Notre Dame, IN 46556}

\author{T. R. O'Brien}
\altaffiliation{Current address: Department of Electrical and Computer Engineering, University of Illinois at Urbana-Champaign, Urbana, IL 61820}
\affiliation{Department of Physics, University of Notre Dame, Notre Dame, IN 46556}

\author{K. J. Schlesinger}
\altaffiliation{Current address: Department of Physics, University of California, Santa Barbara, CA, 93106}
\affiliation{Department of Physics, University of Notre Dame, Notre Dame, IN 46556}

\author{C. D. Dewhurst}
\affiliation{Institut Laue-Langevin, 6 Rue Jules Horowitz, F-38042 Grenoble, France}

\author{L. DeBeer-Schmitt}
\affiliation{Oak Ridge National Laboratory, Oak Ridge, TN 37831-6393}

\author{N. D. Zhigadlo}
\author{J. Karpinski}
\affiliation{Laboratory for Solid State Physics, ETH Zurich, CH-8093 Z\"{u}rich, Switzerland}

\author{M. R. Eskildsen}
\email{eskildsen@nd.edu}
\affiliation{Department of Physics, University of Notre Dame, Notre Dame, IN 46556}

\date{\today}

\begin{abstract}
The vortex lattice (VL) symmetry and orientation in clean type-II superconductors depends sensitively on the host material anisotropy, vortex density and temperature, frequently leading to rich phase diagrams. Typically, a well-ordered VL is taken to imply a ground state configuration for the vortex-vortex interaction. Using neutron scattering we studied the VL in \MgB \ for a number of field-temperature histories, discovering an unprecedented degree of metastability in connection with a known, second-order rotation transition. This allows, for the first time, structural studies of a well-ordered, non-equilibrium VL. While the mechanism responsible for the longevity of the metastable states is not resolved, we speculate it is due to a jamming of VL domains, preventing a rotation to the ground state orientation.
\end{abstract}

\pacs{74.25.Ha, 74.70.Ad, 74.25.Wx, 61.05.fg}

\maketitle

Metastable phases of matter are well-known, with famous examples including supercooling and superheating of liquids and diamond which is one of the many allotropes of carbon. Metastability is almost exclusively observed in connection with first-order transitions, and is often found in frustrated systems where the energy difference between the states is small. The structure of the vortex lattice (VL) in type-II superconductors is known to be highly sensitive to changes in external parameters such as temperature and magnetic field and can therefore naively be expected to display metastability, for example, in connection with first-order transitions such as the VL melting~\cite{RefWorks:865} or the reorientation transition of the rhombic VL found in most superconductors with a four-fold in-plane anisotropy \cite{RefWorks:754}. However, except for phase coexistence in the vicinity of reorientation transitions~\cite{RefWorks:30,RefWorks:576,RefWorks:795,RefWorks:839}, no well-ordered metastable VL phases have been observed to date.

We present extensive small-angle neutron scattering (SANS) studies of the VL in \MgB \ with $\textbf{H} \parallel \textbf{c}$ which is found to exhibit an unprecedented degree of metastability. Previous studies of this compound revealed a continuous, field-driven $30^{\circ}$ VL rotation bounded by two second order transitions. The rotation is understood to arise from a competition between six-fold Fermi surface anisotropies with opposite signs on the $\pi$- and $\sigma$-bands, coupled with the suppression of the $\pi$-band superconductivity by a modest magnetic field~\cite{RefWorks:19,RefWorks:835}. The most striking new results are the existence of highly ordered non-equilibrium VL configurations and the observation of hysteresis in connection with a second order phase transition which we speculate is the result of a jamming-like scenario. As a result, the ground state VL phase diagram is found to differ significantly from our earlier report as well as theoretical predictions~\cite{RefWorks:19,RefWorks:835}.

Small-angle neutron scattering (SANS) experiments were performed at the D22 instrument at Institut Laue-Langevin, the HFIR GP-SANS beam line at Oak Ridge National Laboratory and NG3 at the NIST Center for Neutron Research, using a standard configuration with the applied magnetic field parallel to the incoming neutron beam~\cite{EskildsenFP}. To achieve diffraction the sample and the magnet were rotated and/or tilted together in order to satisfy the Bragg condition for the VL planes. Complete VL diffraction patterns were obtained by adding the scattering, recorded by a 2D position sensitive detector, from several angles. To resolve closely located VL Bragg reflections, a very tight collimation of the neutron beam was used (D22: $0.06^{\circ}$, GP-SANS/NG3: $0.08^{\circ}$~FWHM).

Several different single crystal samples were studied providing consistent results, with only slight variations in the VL transition fields/temperatures. The crystals, with masses in the range 100 -- 200~$\mu$g, were grown using isotopically enriched $^{11}$B to reduce neutron absorption and were comparable to those used in previous SANS experiments by our group~\cite{RefWorks:23,RefWorks:19,RefWorks:12}.

Measurements were performed at temperatures between $T = 2$ and 28~K, and in magnetic fields up to $H = 1.9$~T applied parallel to the crystalline $c$ axis. Several different field-temperature histories were employed: Constant field cooled (FC) from a temperature above $\Tc$ and field ramped (FR) where $H$ was changed at a constant temperature. Other measurements were performed following a damped, small-amplitude field oscillation applied after either a FC or FR procedure.

\begin{figure*}
  \includegraphics{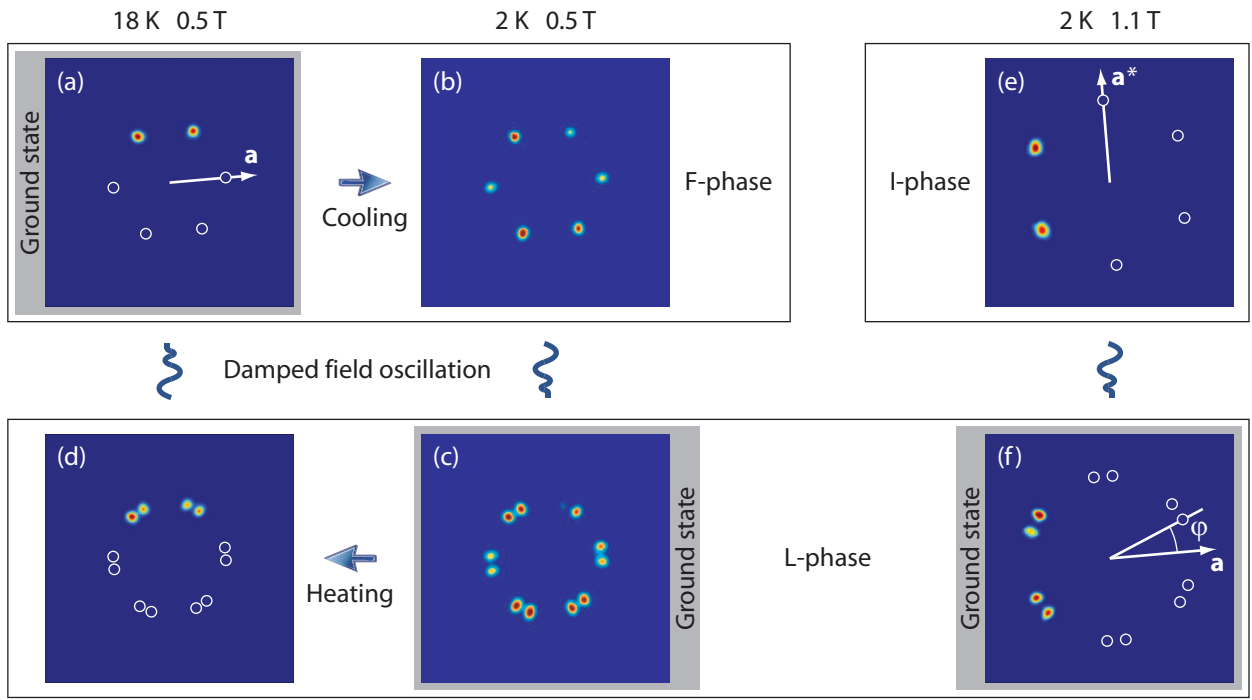}
  \caption{\label{Fig1}
           SANS diffraction patterns showing the different triangular VL phases in \MgB \ with $\textbf{H} \parallel \textbf{c}$. The orientation of the crystalline $a$ axis is shown in panel (a) and the $a^*$ axis in panel (e). The three columns show the VL configurations at 18~K and $0.5$~T (a and d), 2~K and $0.5$~T (b and c), and 2~K and $1.1$~T (e and f). In cases where not all VL reflections were rocked through the Bragg condition, symmetry equivalent positions, obtained by mirroring through and/or rotating around $q = 0$, are indicated by open circles. In the high symmetry F- and I-phases a single-domain VL is observed aligned with Bragg reflections along respectively $\textbf{a}$ and $\textbf{a}^*$. In the intermediate L-phase two VL domains orientations, indicated by $2 \times 6 = 12$ peaks, are observed rotated away from $\textbf{a}$ by an angle $\varphi$ in opposite directions as shown in panel (f). For each field/temperature the ground state VL is obtained following a damped field oscillation with an initial amplitude of 25~mT.}
\end{figure*}

Fig.~\ref{Fig1} shows VL diffraction patterns, obtained at three different places in the mixed state phase diagram of \MgB, which summarize the main result of this report.
Panels (a) and (b) were obtained following a field cooling (FC) procedure from a temperature $T > \Tc$ to 18~K and then 2~K in a constant field of $0.5$~T, and both show VLs oriented with Bragg reflections along the crystalline $a$ axis. In both cases the VL is well-ordered, indicated by very sharp and well-defined Bragg peaks and narrow rocking curves shown in the Supplemental Material. Traditionally such high quality diffraction patterns are taken as an indication that the VL is an equilibrium configuration for the vortex-vortex interaction~\cite{RefWorks:576}. However, applying a small-amplitude field oscillation at 2~K causes a transition to a VL phase which is rotated away from the $a$ axis by an angle $\varphi$, as shown in panel (c). Here a superposition of scattering from two degenerate domain orientations is observed, corresponding to clockwise and anti-clockwise rotations of the VL. The widths of the peaks (radial, azimuthal, longitudinal/rocking curve) stay effectively unchanged. The rotated VL remains robust upon heating to 18~K as shown in (d), but a subsequent field oscillation causes a transition to the $a$ axis aligned phase as in panel (a). Additional measurements showed that in the absence of a field oscillation the rotated VL phase at $0.5$~T (d) persists up to the highest measurable temperature of 28~K.

These results show that the ordered VL in \MgB \ exhibits metastability as it is heated or cooled across the equilibrium phase transition which must exist somewhere between 2 and 18~K. Furthermore, it is necessary to induce vortex motion by a small-amplitude field oscillation in order to drive the VL to the ground state (equilibrium) configuration. A similar effect is observed at $1.1$~T where the field cooled VL is aligned along the $a^*$ axis as shown in panel (e). After a field oscillation, the VL enters the rotated phase (f). In summary, this system presents, for the first time, the opportunity to perform structural studies of a well-ordered, non-ground state VL. Due to the similarity with the tilted hexatic phases in liquid crystals, we adopt the same naming scheme, denoting the VL phases as F, I and L as indicated in Fig.~\ref{Fig1}~\cite{RefWorks:837,Fisch,LCnaming}.

\begin{figure}
  \includegraphics{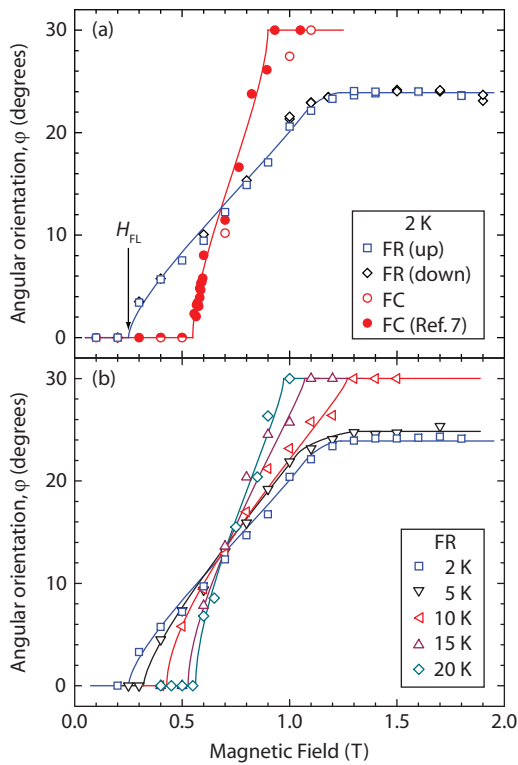}
  \caption{\label{Fig2}(Color online)
           VL orientation angle as a function of applied field.
           Panel (a) shows the the results of measurements at 2~K following a FC procedure and a zero field cooling followed by a FR. The arrow indicates the FL-transition for the ground state VL.
           Panel (b) shows the FR VL orientation at temperatures ranging from 2 to 20~K.
           All lines are guides to the eye.}
\end{figure}

\begin{figure}
  \includegraphics{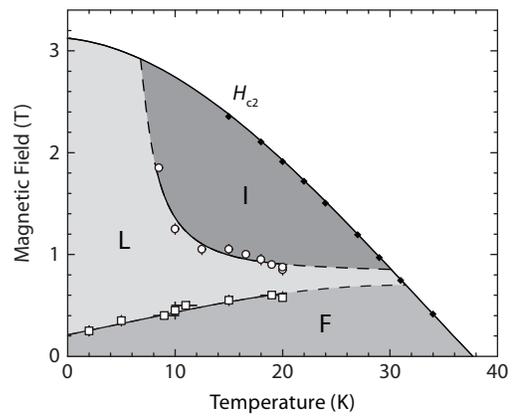}
  \caption{\label{Fig3}
           Ground state VL phase diagram for \MgB \ with $\textbf{H} \parallel \textbf{c}$. The FL-transition is indicated by squares and the LI-transition by circles. The lines are guides to the eye with dashes indicating extrapolations to higher temperatures and fields. Diamonds indicate measured values of $\Hcii$.}
\end{figure}

Fig.~\ref{Fig2}(a) shows a comparison between the VL orientation angle $\varphi$, obtained at 2~K following respectively a FC and a field ramping (FR) procedure where the magnetic field is changed at the measurement temperature.
For fields $\geq 0.2$~T the two procedures result in different VL phases. Furthermore, the ground state VL at this temperature never completes the rotation transition but remains in the L-phase, stabilizing at $\varphi \approx 24^{\circ}$ above $1.2$~T. We note that a FR or a FC followed by a field oscillation both achieve the same result. Fig.~\ref{Fig2}(b) shows the field dependence of the FR VL orientation obtained at temperatures in the range 2 -- 20~K. Only for temperatures of 10~K and above does the ground state VL reach the I-phase ($\varphi = 30^{\circ}$). The ground state VL configuration is summarized in the phase diagram in Fig.~\ref{Fig3}.
Due to decreasing intensity, it was not possible to determine the phase boundaries above $1.8$~T and 20~K. Nonetheless, the constant $\varphi$ for $H \geq 1.3$~T at both 2 and 5~K seen in Fig.~\ref{Fig2} strongly suggests that LI-transition is essentially vertical at low temperatures as indicated by the extrapolation in the phase diagram. Likewise, the fact that the L-phase can be observed following a FC implies that it extends up to $\Hcii$ within a narrow field range.

The ground state VL phase diagram differs markedly from a naive expectation that the effect of the $\pi$-band anisotropy, believed to be responsible for the F- and L-phases, should be largest at low fields (little suppression of $\pi$-band superconductivity) and high temperatures (increased thermal mixing of bands)~\cite{RefWorks:19,RefWorks:835}. Consequently, one would expect the I-phase to occupy the high-field/low-temperature part of the phase diagram in striking contrast to Fig.~\ref{Fig3}. In particular, the reappearance of the L-phase as one follows $\Hcii$ to lower temperatures is unexpected and does not agree with theoretical predictions \cite{RefWorks:835}.

\begin{figure}
  \includegraphics{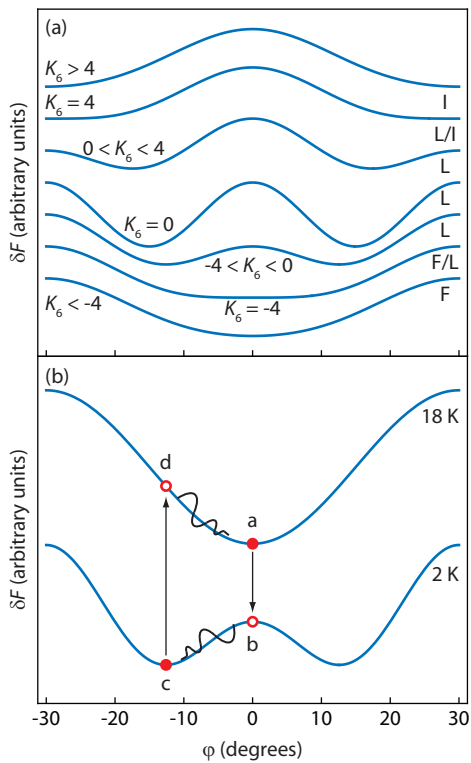}
  \caption{\label{Fig4}(Color online)
           Free energy curves for the VL phases in \MgB.
           Panel (a) shows the progression of free energy curves calculated using Eqn.~(\ref{dF}) with $K_{12} = 1$ and increasing $K_6$. The corresponding ground state VL phase is indicated by each curve. The FL- and LI-transitions occur at $K_6 = \pm 4$.
           Panel (b) shows free energy curves corresponding to the sequence of diffractions patterns at $0.5$~T shown in Fig.~\ref{Fig1}. Solid circles represent ground state VL configurations while open circles represent metastable phases.}
\end{figure}

We now return to the metastable VL phases shown in Fig.~\ref{Fig1}(b), (d) and (e), noting that the existence of such states in connection with a second order phase transition is unexpected as shall be discussed in the following. It should be stressed that the present case is not one of a first order transition broadened by sample inhomogeneities as for example observed in connection with the order-disorder transition associated with the peak effect~\cite{RefWorks:864}. As pointed out by Zhitomirsky and Dao, the continuous rotation of the VL in \MgB \ with second order FL- and LI-transitions shows that the VL free energy as a function of the orientation angle must be given by
\begin{equation}
  \delta F = K_6 \, \cos 6\varphi + K_{12} \, \cos 12\varphi,
  \label{dF}
\end{equation}
with $K_{12} > 0$~\cite{RefWorks:835}. The same free energy expression has been used to model the reorientation transitions observed between tilted hexatic phases in liquid crystals~\cite{RefWorks:836}, where we note no metastable phases are observed. Fig.~\ref{Fig4}(a) shows the evolution of $\delta F$ as $K_6$ is increased from $<-4K_{12}$ to $>4K_{12}$.
In Fig.~\ref{Fig4}(b) we consider the free energy curves corresponding to the cycle of VL phases in Fig.~\ref{Fig1}(a -- d), which shows that the metastable F-phase (b) corresponds to an unstable equilibrium [$d(\delta F)/d\varphi = 0; d^2(\delta F)/d\varphi^2 < 0$] while the metastable L-phase (d) represents a true non-equilibrium situation [$d(\delta F)/d\varphi \neq 0$]. This implies that the VL metastability can not be understood from the single domain free energy thus raising the important question: Which mechanism is responsible for the longevity of the metastable states, preventing them from immediately rotating to the ground state?

A straightforward answer to the above question is that the metastable VLs are held in place by vortex pinning. However, in general metastability is a hallmark of very clean systems, in obvious contrast to such a scenario. More specifically this also seems unlikely given the very low depinning critical current  in clean \MgB \ single crystals similar to the ones used in the present work ($< 2 \times 10^3$~A/cm$^2$ at 5~K and $0.5$~T and decreasing with higher $T$ and $H$) \cite{RefWorks:841}. Moreover, at $0.5$~T the irreversibility line for $\textbf{H} \parallel \textbf{c}$  is in the range 8 -- 14~K \cite{RefWorks:841}, much below the highest measurable temperature of 28~K, where we still observe a metastable L-phase. Finally, the sharp VL Bragg peaks observed in the FC case does not indicate any significant disorder due to vortex pinning 
in contrast to reports for e.g. YNi$_2$B$_2$C \cite{RefWorks:576,RefWorks:795,RefWorks:839}.

We speculate that the mechanism which sustains the metastable configurations is a jamming of counter-rotated VL domains against one another~\cite{RefWorks:866}. In this scenario the VL domains acts as granular entities, preventing each other from rotating to the ground state orientation. If this is the case a close examination of the metastable F- and I-phases should find an onset of domain formation, leading to an azimuthal broadening of the VL peaks which would be most visible at higher fields as the resolution improves with $H$. Such an azimuthal broadening is indeed observed at 2~K and $1.1$~T, increasing from a resolution limited $\sim 4^{\circ}$ in the ground state L-phase to $\sim 7.5^{\circ}$ in the metastable I-phase, as shown in Fig.~S3 and discussed in detail in the Supplemental Material. No broadening is expected or observed in the metastable L-phase. Further studies are needed to fully explore this jamming hypothesis. A key question in this regard will be to investigate the robustness of the VL domain boundaries necessary to prevent rotation. Theoretically, recent work by Deutsch and Shapiro considered domain boundaries in a square VL~\cite{RefWorks:674}. Such studies should be extended to investigate how this affects the VL configuration.

Before concluding, we note that SANS studies of the VL in UPt$_3$ used a temperature quenching technique to infer information about the superconducting A-phase which exist just below $\Hcii$ in this material~\cite{RefWorks:215}. In this case a rotated VL identical to the L-phase was interpreted as originating from a particular order parameter symmetry. Our results on \MgB, which does not have the exotic pairing symmetry of UPt$_3$, are qualitatively similar, suggesting that this interpretation may have to be reconsidered.

In summary, we discovered a high degree of metastability in the \MgB \ VL, which allows measurements in non-equilibrium configurations. The ground-state VL phase diagram was determined and found to differ substantially from theoretical predictions. We speculate that a jamming scenario is responsible for the VL metastability.

We acknowledge discussions with X. Cheng, G. Crawford, R. Cubitt, E. M. Forgan, M. Gingras, M. Konczykowski, C. Olson Reichhardt and C. Reichhardt, and technical assistance at NCNR by J. G. Barker and A. J. Jackson.
This work was supported by the Department of Energy, Basic Energy Sciences under Award No. DE-FG02-10ER46783.
T.R.O'B recognizes support from the Notre Dame Glynn family honors program, the College of Science Undergraduate Research Fund and the ND REU program (National Science Foundation grant PHY05-52843). K.J.S. recognizes support from the Notre Dame Institute for Scolarship in the Liberal Arts.
N.D.Z. and J.K. was supported by the National Center of Competence in Research MaNEP (Materials with Novel Electronic Properties).
The Research at Oak Ridge National Laboratory's High Flux Isotope Reactor was sponsored by the Scientific User Facilities Division, Office of Basic Energy Sciences, U. S. Department of Energy.
We acknowledge the support of the National Institute of Standards and Technology, U.S. Department of Commerce, in providing neutron research facilities used in this work. This work utilized facilities supported in part by the National Science Foundation under Agreement No. DMR-0944772.

\bibliographystyle{apsrev4-1}

\begin{thebibliography}{99}
\bibitem{LCnaming}
  In liquid crystals SmF, SmL and SmI are smectic phases with long range orientational (but not positional) order. While CrH and CrK are used for the
  crystalline phases corresponding to SmF and SmI respectively, there is no equivalent name for the SmL phase. We have therefore chosen to use F, L, I for the
  VL phases in \MgB.
\end{thebibliography}

\newpage

\end{document}


\newcommand{\MgB}{MgB$_2$}
\newcommand{\UPt}{UPt$_3$}
\newcommand{\Tc}{T_{\text{c}}}
\newcommand{\Hci}{H_{\text{c1}}}
\newcommand{\Hcii}{H_{\text{c2}}}
\newcommand{\fq}{\phi_0}
\newcommand{\lamn}{\lambda_n}

\title{Supplemental Material: Metastable Vortex Lattice Phases in Superconducting {\MgB}}

\author{P. Das}
\author{C. Rastovski}
\affiliation{Department of Physics, University of Notre Dame, Notre Dame, IN 46556}

\author{T. R. O'Brien}
\affiliation{Department of Physics, University of Notre Dame, Notre Dame, IN 46556}

\author{K. J. Schlesinger}
\affiliation{Department of Physics, University of Notre Dame, Notre Dame, IN 46556}

\author{C. D. Dewhurst}
\affiliation{Institut Laue-Langevin, 6 Rue Jules Horowitz, F-38042 Grenoble, France}

\author{L. DeBeer-Schmitt}
\affiliation{Oak Ridge National Laboratory, Oak Ridge, TN 37831-6393}

\author{N. D. Zhigadlo}
\author{J. Karpinski}
\affiliation{Laboratory for Solid State Physics, ETH Zurich, CH-8093 Z\"{u}rich, Switzerland}

\author{M. R. Eskildsen}
\affiliation{Department of Physics, University of Notre Dame, Notre Dame, IN 46556}

\date{\today}


\maketitle

Here we present an analysis of the \MgB \ vortex lattice (VL) Bragg peak widths measured by small-angle neutron scattering (SANS) with $\textbf{H} \parallel \textbf{c}$. The measurements were performed following a variety of magnetic field/temperature histories to explore both ground state and metastable VL configurations. Measurements of the ground state phases were performed following a magnetic field ramp at constant temperature with both increasing (FR up) and decreasing (FR down) fields, or following a small-amplitude (25~mT) damped field oscillation (FO) after a temperature change. Metastable VL phases were measured when a temperature change across one of the equilibrium phase transition was {\em not followed} by a FR or FO.

\begin{figure}[b]
  \includegraphics{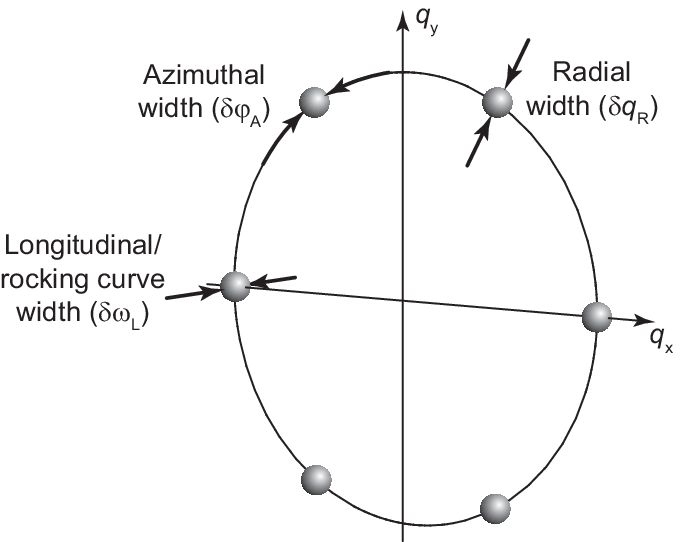}
  \caption{\label{SOMFig1}
           Schematic showing the three different VL widths measurable by SANS.}
\end{figure}

Three widths can be measured, as indicated in Fig.~\ref{SOMFig1}.
The radial ($\delta q_{\text{R}}$) and azimuthal ($\delta \varphi_{\text{A}}$) widths are related to the positional and orientational order of the VL, while the longitudinal curve width ($\delta \omega_{\text{L}}$) is a measure of the straightness of the vortices. The first two widths were obtained by a two-dimensional Gaussian fit to the VL peaks recorded on the detector (Fig.~1 in main paper). The longitudinal width was obtained from so-called rocking curves, which is the intensity variation of the VL Bragg peaks as they are rotated through the Ewald sphere. The rocking curves were corrected for the Lorentz factor (angle at which peaks move through the Ewald sphere) and fitted by a Voigt function. Further details concerning SANS and how the rocking curves are measured can be found in Ref.~\onlinecite{EskildsenThesis}.

The measured widths will be compared to the experimental resolution obtained by the following expressions~\cite{EskildsenThesis}:
\begin{eqnarray}
  \Delta q_{\text{R}}^2 & = & \left( k \, \Delta(2\theta)_{\text{WS}} \right)^2 + \left( k \, \Delta(2\theta)_{\text{BD}} \right)^2 \\
  \Delta \varphi_{\text{A}}^2 & = & \left( k/q \; \Delta(2\theta)_{\text{BD}} \right)^2 \\
  \Delta \omega_{\text{L}}^2 & = & \Delta(2\theta)_{\text{WS}}^2 + \Delta(2\theta)_{\text{BD}}^2.
\end{eqnarray}
Here $k = 2\pi/\lamn$ is the neutron wavevector magnitude and $\lamn$ is the wavelength. The VL scattering vector (for a hexagonal symmetry) is given by $q = 1.075 \times 2\pi (B/\fq)^{1/2}$ where $B$ is the magnetic field and $\fq = h/2e = 2067$~Tnm$^2$ is the flux quantum. There are two contributions to the resolution. The first comes from the wavelength spread $\Delta \lamn/\lamn = 10\%$ and is given by
\begin{equation}
  \Delta(2\theta)_{\text{WS}} = \frac{q \lamn}{2\pi} \; \sqrt{\frac{4 \ln 2}{3}} \; \frac{\Delta \lamn}{\lamn},
\end{equation}
where the numerical factor is due to the conversion from variance to FWHM. The second contribution ($\Delta(2\theta)_{\text{BD}}$) is from the beam divergence, which was determined from the size of the undiffracted beam spot on the detector. Since $k \gg q$ the longitudinal resolution is substantially higher than the in-plane resolution.

\begin{figure}
  \includegraphics{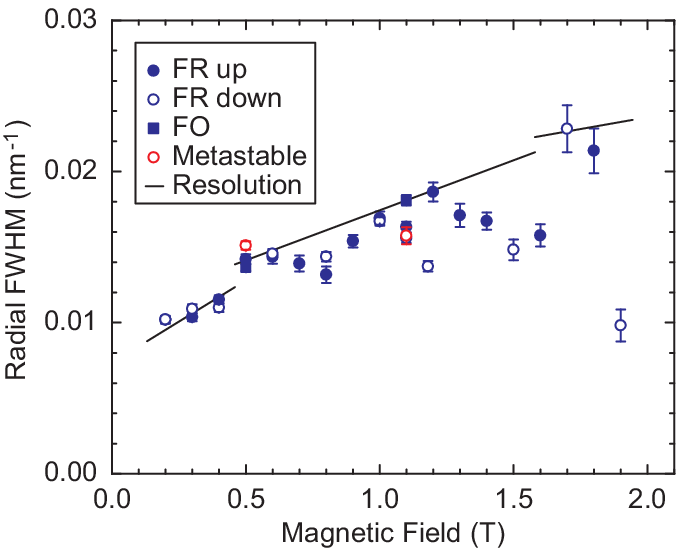}
  \caption{\label{SOMFig2}
           Field dependence of the radial width ($\delta_{\text{R}}$) of the VL Bragg peaks at 2~K for different field/temperature histories: Increasing field ramp (FR up), decreasing field ramp (FR down), field oscillation (FO) and cooling/heating in constant field (Metastable). The solid line is the experimental resolution with the discontinuities corresponding to changes of the neutron wavelength ($\lamn = 10$, 7 and 5~nm with increasing field).}
\end{figure}

Fig.~\ref{SOMFig2} shows the field dependence of the radial width  at 2~K.
This is found to increase with field, but remains resolution limited throughout the range of measurements. In particular there is no measurable difference between ground state and metastable VLs, showing that within the measurement precision the variation of the lattice plane spacing at all fields is the same for both phases. The calculated experimental resolution is possibly slightly overestimated. The measured widths at high field, which fall substantially below the resolution, can be attributed to the rapidly decreasing scattering intensity and resulting poor fits which yield unphysical results. The radial width can, in principle, be used to estimate the size of the VL domains in the direction perpendicular to the field. However, since this width is resolution limited ($\delta q_{\text{R}} \approx \Delta q_{\text{R}}$) it only provides a lower bound $\zeta_{\text{R}} \geq 2/\delta q_{\text{R}} \approx 100 - 200$~nm.

\begin{figure}
  \includegraphics{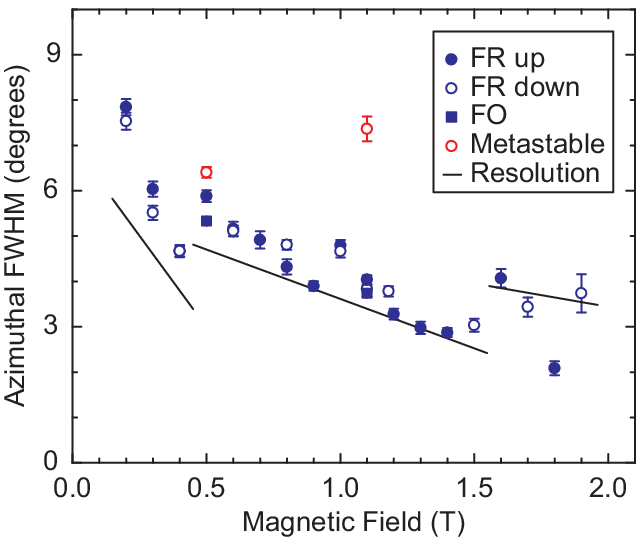}
  \caption{\label{SOMFig3}
           Field dependence of the azimuthal width ($\delta \varphi_{\text{A}}$) of the VL Bragg peaks at 2~K. The solid line is the experimental resolution with the discontinuities corresponding to changes of the neutron wavelength.}
\end{figure}

We now turn to the field dependence of the azimuthal width shown in Fig.~\ref{SOMFig3}.
Here all ground state VLs are again found to be resolution limited, whereas the mestable VL at 1~T is found to be significantly broadened. Similarly the metastable VL at $0.5$~T is broader than the ground state but in this case this is within the statistical fluctuations of the data. For the ground state we find a lower bound of the VL domain size of $\zeta_{\text{A}} \geq 2/(q \, \delta \varphi_{\text{A}}) \approx 150 - 300$~nm in good agreement with the estimate or $\zeta_{\text{R}}$ above. We interpret the broadening of the metastable VL at $1.1$~T as being due to a small rotation (onset of transition to L-phase) which is insufficient to clearly separate the peak into two.

\begin{figure}[t]
  \includegraphics{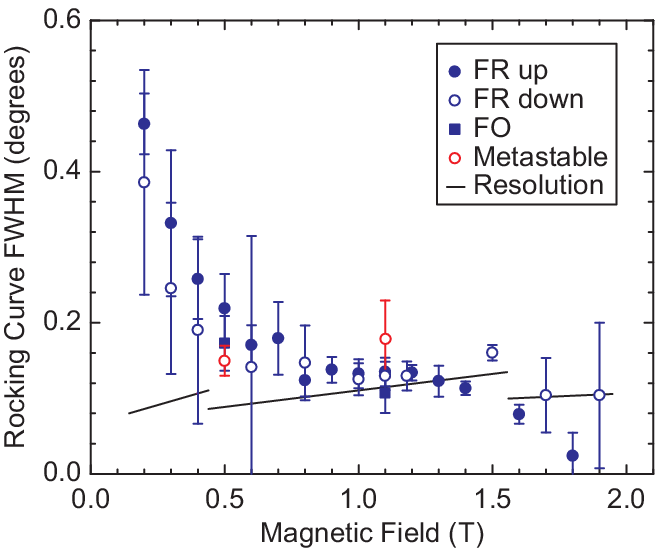}
  \caption{\label{SOMFig4}
           Field dependence of the rocking curve width ($\delta \omega_{\text{L}}$) of the VL Bragg peaks at 2~K. The solid line is the experimental resolution with the discontinuities corresponding to changes of the neutron wavelength.}
\end{figure}

The field dependence of the VL rocking curve width is given in Fig.~\ref{SOMFig4}.
This show a conventional monotonic decrease with increasing field, indicative of a straightening of the vortices as they are pushed closer together and their mutual repulsion become stronger. At roughly 1~T the rocking curve width becomes resolution limited. An estimate of the length of straight vortex segments is given by $\zeta_{\text{L}} \geq 2/(q \, \delta \omega_{\text{L}})$ which varies from 4 to 6~$\mu$m. No difference between metastable and ground state VL configurations is observed.

\begin{figure}
  \includegraphics{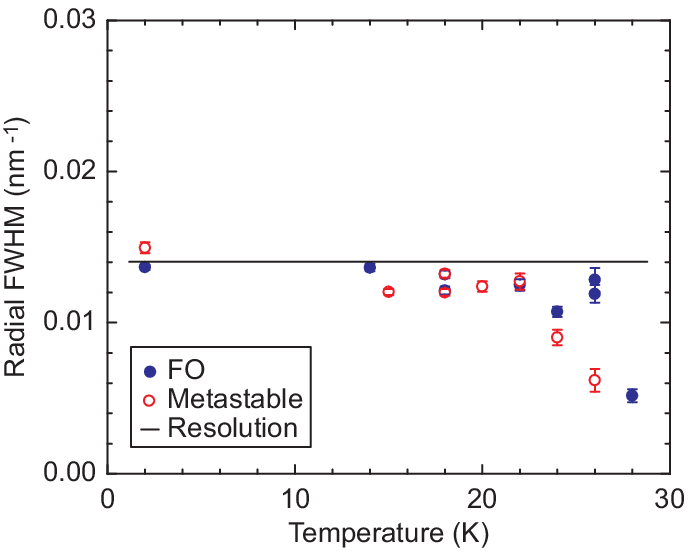}
  \caption{\label{SOMFig5}
           Temperature dependence of the radial width of the Bragg peaks at $0.5$~T in the ground state (FO) and metastable VL phases. The solid line is the calculated experimental resolution for $\lamn = 7$~nm.}
\end{figure}

\begin{figure}
  \includegraphics{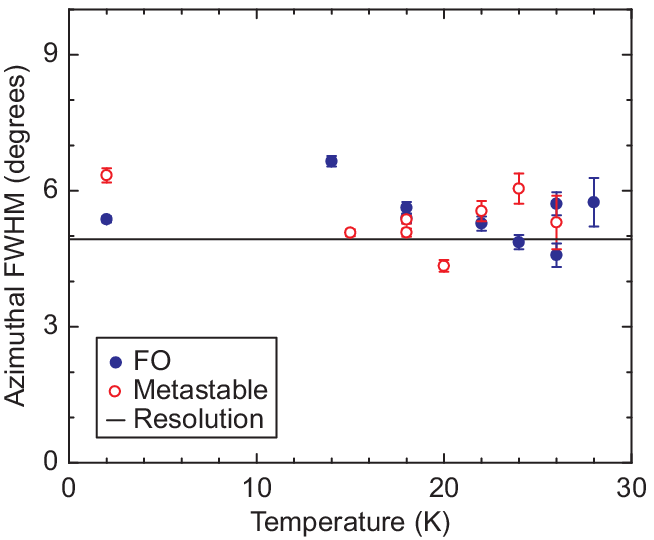}
  \caption{\label{SOMFig6}
           Temperature dependence of the azimuthal width of the VL Bragg peaks at $0.5$~T. The solid line is the experimental resolution.}
\end{figure}

\begin{figure}
  \includegraphics{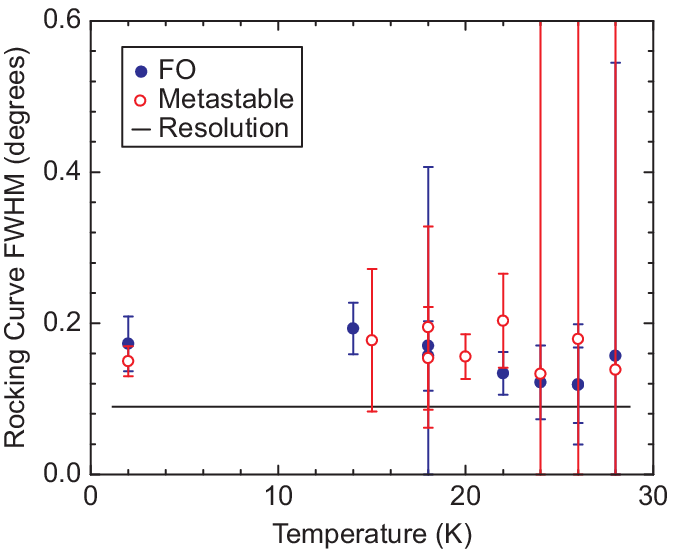}
  \caption{\label{SOMFig7}
           Temperature dependence of the rocking curve width of the VL Bragg peaks at $0.5$~T. The solid line is the experimental resolution.}
\end{figure}

Figs.~\ref{SOMFig5} to \ref{SOMFig7} show the temperature dependence at $0.5$~T of the radial, azimuthal and rocking curve widths respectively.
In all cases no temperature dependence is observed within the measurement precision. Similarly, no differences between metastable and ground state VLs are found. In the case of the radial width we again find this to be resolution limited, with fits at high temperature which fall below the experimental resolution. The low intensity at high temperature also leads to very large errors on the rocking curve widths as seen in Fig.~\ref{SOMFig7}.


\bibliographystyle{apsrev4-1}
\bibliography{Bib}
